\documentclass{ws-ijmpcs}
\def\trimmarks{}

\usepackage{graphicx}
\usepackage{lineno}

\begin{document}

\renewcommand{\thefootnote}{\alph{footnote}}


%
\catchline{}{}{}{}{}
%

\title{Studies of Transverse-Momentum-Dependent distributions with A~Fixed-Target~ExpeRiment using the LHC beams (AFTER@LHC)}

\author{L.~Massacrier$^{1,2,}$\footnote{contact e-mail: massacri@lal.in2p3.fr} , M.~Anselmino$^{3}$, R.~Arnaldi$^{3}$, S.~J.~Brodsky$^{4}$, V.~Chambert$^{2}$, W.~den~Dunnen$^5$, J.P.~Didelez$^{2}$, B.~Genolini$^{2}$, E.G.~Ferreiro$^{6}$, F.~Fleuret$^{7}$, Y.~Gao$^{8}$, C.~Hadjidakis$^{2}$, I.~H\v{r}ivn\'{a}\v{c}ov\'{a}$^{2}$, J.P.~Lansberg$^{2}$, C.~Lorc\'e$^{4,9}$, R.~Mikkelsen$^{10}$, C.~Pisano$^{11}$, A.~Rakotozafindrabe$^{12}$, P.~Rosier$^{2}$, I.~Schienbein$^{13}$, M.~Schlegel$^{5}$, E.~Scomparin$^{4}$, B.~Trzeciak$^{14}$, U.I.~Uggerh$\o$j$^{10}$, R.~Ulrich$^{15}$, and Z.~Yang$^{8}$.}

\address{$^{1}$ LAL, Universit\'e Paris-Sud, CNRS/IN2P3, Orsay, France \\
$^{2}$ IPNO, Universit\'e Paris-Sud, CNRS/IN2P3, F-91406, Orsay, France \\
$^{3}$ Dip. di Fisica and INFN Sez. Torino, Via P. Giuria 1, I-10125, Torino, Italy \\ 
$^{4}$ SLAC Nat. Accel. Lab., Theoretical Physics, Stanford U. Menlo Park, CA 94025, USA \\
$^{5}$ Institute for Theoretical Physics, T\"ubingen U., D-72076 T\"ubingen, Germany \\
$^{6}$ Dpt. de F\'isica de Part\'iculas, Universidade de Santiago de C., 15782 Santiago de C., Spain \\
$^{7}$ Laboratoire Leprince Ringuet, \'Ecole Polytechnique, CNRS/IN2P3, 91128 Palaiseau, France \\
$^{8}$ CHEP, Department of Engineering Physics, Tsinghua University, Beijing, China\\
$^{9}$ IFPA, AGO Dept., Universit\'e de Li\`ege, Sart-Tilman, 4000 Li\`ege, Belgium\\
$^{10}$ Dept. of Physics and Astronomy, University of Aarhus, Denmark\\
$^{11}$ Nikhef \& Dept. of Phys. \& Astr., VU Amsterdam, NL-1081 HV Amsterdam, The Netherlands\\
$^{12}$ IRFU/SPhN, CEA Saclay, 91191 Gif-sur-Yvette Cedex, France\\
$^{13}$ LPSC, Univerist\'e Joseph Fourier, CNRS/IN2P3/INPG, F-38026 Grenoble, France\\
$^{14}$ FNSPE, Czech Technical U., Prague, Czech Republic\\
$^{15}$ Institut f\"ur Kernphysik, Karlsruhe Institute of Technology (KIT), 76021 Karlsruhe, Germany
}



\maketitle


\begin{abstract}
We report on the studies of Transverse-Momentum-Dependent distributions (TMDs)
at a future fixed-target experiment --AFTER@LHC-- using the $p^+$ or Pb ion LHC 
beams, which would be the most energetic fixed-target experiment ever performed. 
AFTER@LHC opens new domains of particle and nuclear physics by complementing 
collider-mode experiments, in particular those of RHIC and the EIC projects. 
Both with an extracted beam by  a bent crystal or with an internal gas target, the 
luminosity achieved by AFTER@LHC surpasses that of RHIC 
by up to 3 orders of magnitude. With an unpolarised target, it
allows for measurements of TMDs such as the Boer-Mulders quark distributions and the 
distribution of unpolarised and linearly polarised gluons in unpolarised protons. 
Using polarised targets, one can access the quark and gluon Sivers TMDs
through  single transverse-spin asymmetries in
Drell-Yan and quarkonium production. In terms of kinematics, the fixed-target mode 
combined with a detector covering $\eta_{\rm lab} \in [1,5]$
allows one to measure these asymmetries at large $x^\uparrow$ in the polarised nucleon. 
\keywords{TMD; AFTER@LHC; single spin asymmetry; Drell-Yan; quarkonium}
\end{abstract}

\ccode{PACS numbers:07.90.+c, 21.10.Hw, 13.20.Gd}

\section{The AFTER@LHC project}

AFTER@LHC\cite{Brodsky:2012vg,Lansberg:2012kf,Lansberg:2012wj,Rakotozafindrabe:2013cmt,Rakotozafindrabe:2013au} is a proposal of a future multi-purpose fixed-target experiment (FTE) using the multi-TeV proton or heavy ion beams of the LHC. Such a proposal allows for an exceptional testing ground for QCD at unprecedented laboratory energies and momentum transfers. Important advances in nuclear and hadron physics in the past decades were achieved thanks to FTE: particle discoveries ($\Omega^{-}$(sss), J/$\psi$, $\Upsilon$), first evidence of the creation of a quark-gluon plasma in heavy-ion collisions, discovery of anomalously large single- and double-spin\cite{Adams:1991cs,Crosbie:1980tb} correlations in hadron-hadron collisions, etc. Indeed, the fixed-target mode offers several advantages with respect to the collider mode:
\begin{itemlist}
\item{Outstanding luminosities are obtained thanks to the target high density;}
\item{The far backward region in the center-of-mass system (c.m.s), corresponding to an acceptance of 1 $\leq \eta_{\rm lab} \leq$ 5, is not limited by geometrical constraints;}
\item{A large number of target species can be studied; and}
\item{The c.m.s energy is the same for $pp$, $pd$, $pA$ collisions (115 GeV with a proton beam of 7 TeV), and for Pb$p$, PbA (72 GeV with a Pb beam of 2.76 TeV).}
\end{itemlist}

The objective of performing FTE with the $p^{+}$ and Pb LHC beams is threefold. First,
one wishes to significantly advance our understanding of the large-$x$
gluon, antiquark  and  heavy-quark content
in the nucleon and nucleus. There are many motivations for this:
\begin{itemlist}
\item The uncertainties of state-of-the-art PDF fits are very large for
$x \gtrsim 0.5$. This could be crucial 
 to confirm  possible observed excesses in the data at the LHC or Future Circular Collider and
to characterise possible Beyond-the-Standard-Model discoveries at these facilities;
\item It is equally important for high-energy neutrino and cosmic-rays
physics to better constraint the charm-content of the nucleon;
\item In the nuclear case, the  EMC effect\cite{Aubert:1983xm} is still an open problem more
than two
decades after its discovery. The search for a possible \textit{gluon} EMC effect
is essential to understand its origin and the connections
with short-range correlations in nuclei; 
\item The understanding of the initial state of heavy-ion collisions,
and thus of nuclear PDFs,  is also crucial to study the deconfinement
of quark and gluons at RHIC and the LHC;
\item To further our tests of QCD, the
search and study of rare proton fluctuations where
one gluon carries most of the proton momentum is extremely appealing.
\end{itemlist}

Second, one wishes to  advance our understanding  of the dynamics and
the spin of \textit{gluons} inside polarised and unpolarised nucleons.
This is motivated such as:
\begin{itemlist}
\item Our understanding of the spin of the nucleon made of partons is still
incomplete. A possible missing contribution could arise from their angular momentum.
\item It is important to study spin-dependent object, which can also be defined for unpolarised nucleons. One example of these is the
distribution of linearly-polarised gluons in unpolarised protons. Once
these are known, any hadron collider can in principle be used
to do spin physics.
\item On the way of studying the parton angular momentum, via transverse-momentum-dependent observables for instance, one would also
test fundamental properties of QCD such as the factorisation or
universality of initial- and final-state radiations.
\end{itemlist}

Third, one wishes to make a decisive step forward in the study of
heavy-ion collisions at ultra-relativisitic energies with measurements
towards large rapidities where one of the colliding nuclei is nearly at
rest. This is motivated as such:
\begin{itemlist}
\item In order to better explore the time evolution and longitudinal
expansion of the deconfined matter,  it is essential to measure new hard
probes
in the wide longitudinal-momentum range accessible in the fixed-target mode.
\item Hard probes studies with the same experiment in asymmetric
heavy-ion collisions and in proton-nucleus can provide
key insights on the factorisation of cold nuclear matter effect (CNM) in the environment
of two heavy ions. If, in some corners of the phase space, such
a factorisation is broken, the subtraction of these CNM could simply be impossible.
\item In order to use azimutal-asymmetry measurements as a tool to
study the properties of the deconfined matter, it is essential to
understand their
origin. By measuring them up to large rapidities, one can put more
stringent constraints on  how they are formed, from hydrodynamical origin
or from initial-state radiations.
\end{itemlist}

\section{Beam extraction with a bent crystal vs. an internal gas target}

Two different technological options are currently under investigation to make the highly energetic LHC beams colliding onto a target. First, a bent crystal is positioned in the halo of the LHC beam such that a few protons (or Pb ions) per bunch per pass would be channelled in the lattice of the crystal, and following its curvature, would be deflected by few mrad w.r.t the beam axis. Such an extraction technique is a convenient, efficient and cost-effective way to obtain a clean and well collimated beam, without affecting the LHC performances. This technology was successfully tested for protons at the SPS\cite{Arduini:1997kh}, Fermilab\cite{Asseev:1997yi}, Protvino\cite{Afonin:2012zz} and for Pb ions at the SPS\cite{Scandale:2011za}. It was proposed as a smart alternative for the upgrade of the LHC collimation system and will be tested by the LUA9 Collaboration\cite{LHCC2011} with the 7 TeV LHC beam, at IR7 after Long Shutdown 1. With a bent crystal, one expects to extract an average of 15 protons each 25 ns (i.e. 5 $\times$ $10^{8}$ $p^{+}$s$^{-1}$) from the LHC-beam losses and about 2 $\times$ $10^{5}$ Pb s$^{-1}$. It has been shown\cite{Baurichter:2000wk} that one can expect a degradation of the crystal at the level of 6$\%$ per $10^{20}$ particles/cm$^{2}$ (about 1 year of operation). To cope with such a degradation, the crystal has to be moved by less than a millimeter each year so that the beam halo hits an intact spot of the crystal. This operation can be repeated almost at will. 
In Tab. 1, the instantaneous and yearly luminosities (assuming 10$^7$s of $p^{+}$ beam and 10$^{6}$s of Pb beam per year) are reported for $p^{+}$ and Pb beams on targets of various thicknesses. Integrated luminosities as large as 20 fb$^{-1}$ are reached with a 1m-long target of liquid hydrogen, which is as large as the data sample collected at 7 and 8 TeV at the LHC. \newline
Second, the LHC beam goes through an internal gas target installed in one of the existing LHC experiments or in a new one. Such an internal gas target option is currently tested by the LHCb collaboration via a luminosity monitor\cite{LHCb:SMOG,FerroLuzzi:2005em,LHCb:2014} (SMOG). A pilot run of $p^{+}$ beam (Pb beam) on a Neon gas target was successfully performed in 2012 (2013) at a c.m.s energy of $\sqrt{s_{NN}}$ = 87 GeV (54 GeV). SMOG was tested for few hours only in a row during data taking, with non-getterable gases. No decrease of the LHC performances was observed. More studies are needed to confirm that the internal gas target system can be run over extended periods of time, without any interferences on other LHC experiments and to check the behaviour of the gas (e.g. the maximal pressure that can be reached). Assuming a gas pressure\footnote{We remind that the LHC "vacuum" pressure is $10^{-12}$ bar.} of $10^{-9}$ bar, one can calculate the instantaneous luminosity as follows: $\cal{L} = \phi_{\rm{beam}} \times (\frac{N_{A}}{\rm{22400}} \times \rm{P} \times \ell)$ where $\phi_{\rm{beam}}$ is the $p^{+}$ (or Pb) flux in the LHC, P the gas pressure, and $\ell$ the usable gas zone. In the case of the $p^{+}$ beam, $\phi_{\rm beam}$ = 3.14 $\times 10^{18}$ $p^{+}$ s$^{-1}$ and for the Pb beam, $\phi_{\rm beam}$ = 4.6 $\times 10^{14}$ Pb s$^{-1}$. Instantaneous and yearly luminosites expected with an internal gas target are reported in Tab. 1. Provided that the runs can last as long as a year in both cases, luminosities in $pA$ are similar for the bent crystal and for the internal gas target scenario. However in $pp$, in order to get luminosities as large as 10 fb$^{-1}$ yr$^{-1}$ with the internal gas target, a pressure of $10^{-7}$ bar is required for the gas, which is challenging. 
\begin{table}
\begin{center}
\tbl{Expected luminosities obtained for a 7 (2.76) TeV proton (Pb) beam extracted by means of bent crystal and obtained with an internal gas target.}
{\begin{tabular}{|c c c c c c c|}
  \hline
  Beam & Target & Thickness & $\rho$ & A & $\cal{L}$ & $\int{\cal{L}}$  \\ 
   &  & (cm) & (g.cm$^{-3}$) &  & ($\mu$b$^{-1}$.s$^{-1}$) & (pb$^{-1}$.y$^{-1})$  \\ \hline
  p & Liquid H & 100 & 0.068 & 1 & 2000 & 20000 \\
  p & Liquid D & 100 & 0.16 & 2 & 2400 & 24000 \\
  p & Pb & 1 & 11.35 & 207 & 16 & 160 \\ \hline 
  Pb & Liquid H & 100 & 0.068 & 1 & 0.8 & 0.8 \\
  Pb & Liquid D & 100 & 0.16 & 2 & 1 & 1 \\
  Pb & Pb & 1 & 11.35 & 207 & 0.007 & 0.007 \\ \hline \hline
  Beam & Target & Usable gas zone & Pressure &  & $\cal{L}$ & $\int{\cal{L}}$    \\
       &  & (cm) & (Bar) &  & ($\mu$b$^{-1}$.s$^{-1}$) & (pb$^{-1}$.y$^{-1})$  \\ \hline
    p  & perfect gas & 100 & $10^{-9}$ & & 10 & 100 \\ \hline
    Pb & perfect gas & 100 & $10^{-9}$ & & 0.001 & 0.001 \\ 
       
  \hline
\end{tabular}}
\end{center}
\end{table}
In both scenari, it is technologically possible to polarize the target. For an overview of target polarization techniques see Ref. \refcite{Target:2014}. For the bent-crystal case, the main constraint in the choice of the target polarization technology is the space available in the underground LHC complex, restricting the possibilities to two choices: continuous Dynamic nuclear Polarisation (DNP), or a HD target\cite{Didelez:1995hm}. Both are more compact technologies than the frozen-spin one. CERN has expertise in DNP technology using NH$_{3}$ or Li$_{6}$D materials\cite{Berlin:2011}. Only two groups worldwide are specialized in HD targets: one at TJNAF (USA) and one at RCNP (Japan). Concerning the internal gas target solution, atomic beam source technology, similarly to what is done in HERMES or optical pumping as in SLAC are candidate technologies to polarize the target\cite{Target:2014}. One should mention that the fraction of polarizable nuclei over the total number of nuclei is generally larger for polarized gas target.

\section{TMD studies}

\subsection{Access to the distribution of linearly-polarized gluons in unpolarized protons}

The distribution of linearly-polarized gluons in unpolarized protons is encoded in $h_{1}^{\perp g}$($x$,$k_{T}$,$\mu$) and can be studied without the need of polarizing the target. Such an effect (known as "Boer-Mulders" effect for the quark case\cite{Boer:1997nt}) arises from the correlation between the gluon $k_{T}$ and its spin. The low-$p_{T}$ spectra of scalar and pseudo-scalar quarkonia ($\chi_{c0}$, $\chi_{b0}$, $\eta_{c}$, $ \eta_{b}$) are affected differently\cite{Boer:2012bt} by the linear polarisation of the gluons. Thanks to the large boost ($\gamma$ $\simeq$ 60) in the fixed-target mode, AFTER@LHC could access the low-$p_{T}$ C-even quarkonium. The measurement of the $\eta_{c}$ production is a good candidate for such a kind of studies. Recently, the LHCb Collaboration did the first measurement\cite{Aaij:2014bga} of the $\eta_{c}$ hadroproduction, in the $p\bar{p}$ decay channel, for $p_{T}$ greater than 6 GeV/c. More studies are needed to confirm the feasibility of this measurement at low-$p_{T}$ with AFTER@LHC. \newline
It has also been proposed\cite{Dunnen:2014eta} that the distribution of linearly-polarized gluons in unpolarized protons could be extracted from the measurement of a quarkonium with a back-to-back isolated photon. It has been shown\cite{Lansberg:2014myg} that this observable is still sensitive to gluons at large $x$ at AFTER@LHC energies. One can expect a differential cross section of the order of tens of fb/GeV. J/$\psi$-pair production is also interesting for spin related studies with unpolarised protons.

\subsection{Single Transverse Spin Asymmetries with polarized protons}

With a polarized target, one can access the Sivers functions (encoding the correlation between the proton spin and the parton angular momentum) for the quarks and gluons. 
The existence of a nonzero gluon Sivers effect can be probed by studying single transverse spin assymetries (STSA) in $\eta_{Q}$ production, especially at low-$p_{T}$, which is a clean gluon-sensitive probe. Quarkonia, such as J/$\psi$ and $\Upsilon$, are also good probes to look for the gluon Sivers effect. High precision data are needed for such a kind of studies, and the high luminosities reached by AFTER@LHC would definitively help perform STSA studies. \newline
The quark Sivers effect can be probed with the Drell-Yan (DY) process. AFTER@LHC is competitive with DY measurements to be performed at COMPASS\cite{Quintans:2011zz}, Fermilab\cite{Isenhower:2012vh} at large $x^{\uparrow}$, and with the proposal P1039\cite{Brown:2014sea} at low $x^{\uparrow}$.
An asymmetry up to 10$\%$ has been predicted\cite{Ansel:2013} at AFTER@LHC for DY in the backward region, $x_{F} < 0$. These studies are crucial to test QCD by determining if the quark Sivers function changes sign between Semi-Inclusive DIS and DY pair production. \\
Finally, we stress the possibility to study STSA in $p^{\uparrow}$A collisions using the Pb beam. 

\section{Conclusion}

AFTER@LHC provides a novel testing ground for QCD. High luminosities are achievable in $pp$ and $pA$ at $\sqrt{s_{NN}}$ = 115 GeV as well as Pbp and PbA at $\sqrt{s_{NN}}$ = 72 GeV. TMD studies can be performed with and without polarizing the target, thanks to e.g the study of low-$p_{T}$ quarkonia production. First fast simulations performed with a LHCb-like setup are really promising.  \newline

\footnotesize

\textbf{Acknowledgments.}
This research was supported in part by the French P2IO Excellence Laboratory, the French CNRS via the grants PICS-06149 Torino-IPNO,
FCPPL-Quarkonium4AFTER \& PEPS4AFTER2 and by the Department of Energy, contract DE-AC02-76SF00515.

\appendix


\begin{thebibliography}{000} 
\footnotesize
\bibitem{Brodsky:2012vg}
 S.J.~Brodsky, F.~Fleuret, C.~Hadjidakis, J.P.~Lansberg,
 {\it  Phys.\ Rept.\ }  {\bf 522} (2013) 239

\bibitem{Lansberg:2012kf}
 J.P.~Lansberg, \textit{et al.}, 
 {\it Few Body Syst.\ } {\bf 53} (2012) 11
 
\bibitem{Lansberg:2012wj}
 J.~P.~Lansberg, {\it et al.},
 {\it PoS QNP} {\bf 2012} (2012) 049

\bibitem{Rakotozafindrabe:2013cmt}
 A.~Rakotozafindrabe, {\it et al.},
 {\it PoS DIS} {\bf 2013} (2013) 250


\bibitem{Rakotozafindrabe:2013au}
 A.~Rakotozafindrabe, {\it et al.},
 {\it Phys.\ Part.\ Nucl.\ }  {\bf 45} (2014) 336

\bibitem{Adams:1991cs}
 D.~L.~Adams {\it et al.} 
{\it  Phys.\ Lett.\ B} {\bf 264} (1991) 462.
%
%
\bibitem{Crosbie:1980tb}
 E.~A.~Crosbie, {\it et al.}, 
{\it Phys.\ Rev.\ D} {\bf 23} (1981) 600.
 
\bibitem{Aubert:1983xm}
   J.~J.~Aubert {\it et al.}  [European Muon Collaboration],
   {\it Phys.\ Lett.\ B} {\bf 123} (1983) 275.


\bibitem{Arduini:1997kh}
G. Arduini, {\it et al.},
{\it Phys. \ Lett. \ B} {\bf 422} (1998) 325.
\bibitem{Asseev:1997yi}
 A.~Asseev, {\it et al.}, 
 {\it Phys.\ Rev.\ ST Accel.\ Beams} {\bf 1} (1998) 022801.

\bibitem{Afonin:2012zz}
 A.~G.~Afonin, {\it et al.}, 
 {\it Phys.\ Rev.\ ST Accel.\ Beams} {\bf 15} (2012) 081001.

\bibitem{Scandale:2011za}
 W.~Scandale, {\it et al.}, 
{\it Phys.\ Lett.\ B} {\bf 703} (2011) 547.

\bibitem{LHCC2011} LHC Committee, {\it minutes of the 107th meeting}, CERN/LHCC 2011-010


\bibitem{Baurichter:2000wk}
 A.~Baurichter, \textit{et al.},
 {\it Nucl.\ Instrum.\ Meth.\ B} {\bf 164-165} (2000) 27.


\bibitem{Hab:1988} W. Meyer, Habilitation thesis, Bonn, Germany (1988).

\bibitem{LHCb:SMOG} C. Barschel, PhD thesis, RWTH Aachen U., Germany, CERN-THESIS-2013-301 (2014)

\bibitem{FerroLuzzi:2005em}
   M.~Ferro-Luzzi,
   {\it Nucl.\ Instrum.\ Meth.\ A} {\bf 553} (2005) 388.
   
\bibitem{LHCb:2014}
R. Aaij \textit{et al.}, [LHCb collaboration], 2014 \textit{JINST} 9 P12005.   

\bibitem{Target:2014} N. Doshita, {\it Talk at AFTER@LHC week, CERN November 2014} 

\bibitem{Didelez:1995hm} J.P. Didelez, 
{\it Nucl. Phys. News }{\bf 4}, (1994) 10

\bibitem{Berlin:2011} A. Berlin \textit{et al.}, 
{\it Procs. of 
    PSTP 2011, St Petersburg, Russia}, 
p.  131.


\bibitem{kohri} H. Kohri \textit{et al.}, 
{\it  Procs. of 
    PSTP 2011, St Petersburg, Russia}, 
p.  142.


\bibitem{Boer:1997nt}
 D.~Boer and P.~J.~Mulders,
 {\it Phys.\ Rev.\ D} {\bf 57} (1998) 5780


\bibitem{Boer:2012bt}
 D.~Boer and C.~Pisano,
 {\it Phys.\ Rev.\ D} {\bf 86} (2012) 094007

\bibitem{Aaij:2014bga}
 R.~Aaij {\it et al.}  [LHCb Collaboration],
 arXiv:1409.3612 [hep-ex].

\bibitem{Dunnen:2014eta}
 W.~J.~den Dunnen, \textit{et al.},
 {\it Phys.\ Rev.\ Lett.}  {\bf 112} (2014) 212001

%
%
\bibitem{Lansberg:2014myg}
 J.~P.~Lansberg, {\it et al.} 
{\it EPJ Web of Conf.} {\bf 85}, 02038 (2015)

\bibitem{Ansel:2013}
M. Anselmino, U. D'Alesio, {\it Talk at "AFTER@ECT$^*$, Trento, February 2013} 

\bibitem{Quintans:2011zz}
 C.~Quintans [COMPASS Collaboration],
{\it  J.\ Phys.\ Conf.\ Ser.\ } {\bf 295} (2011) 012163.


\bibitem{Isenhower:2012vh}
 L.~D.~Isenhower, {\it et al.}, 
 {\it FERMILAB-PROPOSAL-1027}.



\bibitem{Brown:2014sea}
 C.~Brown, {\it et al.}, 
{\it  FERMILAB-PROPOSAL-1039}.

\end{thebibliography}
\end{document}